\documentclass[a4paper,12pt]{article}

\usepackage[left=2.5cm,right=2.5cm,top=2.5cm,bottom=2.5cm]{geometry}
\usepackage{graphicx}
\usepackage{amssymb}
\usepackage{multirow}
\usepackage{amsmath}
\usepackage{psfrag}
\usepackage{setspace}
\usepackage{subcaption}
\usepackage[colorlinks=true,bookmarks=true]{hyperref}
\usepackage{float}
\usepackage{cite}
\usepackage[utf8]{inputenc}
\usepackage{leftidx}

\hypersetup{citecolor=blue}
\newcommand{\dif}{\mathrm{d}}
\newcommand{\Eqref}[1]{(\ref{#1})}
\newcommand{\half}{\frac{1}{2}}
\newcommand{\expo}[1]{\mathrm{e}^{#1}}
\newcommand{\brac}[1]{\left(#1 \right)}
\newcommand{\sbrac}[1]{\left[#1\right]}
\newcommand{\im}{\mathrm{i}}
\newcommand{\Vcal}{\mathcal{V}}

\begin{document}

\title{Charged dilaton black hole with multiple Liouville potentials and gauge fields}
\author{Yen-Kheng Lim\footnote{E-mail: yenkheng.lim@gmail.com}\\\textit{\normalsize{Department of Mathematics, Xiamen University Malaysia,}}\\\textit{\normalsize{43900 Sepang, Malaysia}}}
\date{\normalsize{\today}}
\maketitle

\begin{abstract}
  A solution to an Einstein--Maxwell--dilaton-type theory with $M$ Liouville potentials and $N$ gauge fields is presented, where $M$ and $N$ are arbitrary integers. This exact solution interpolates between the Lifshitz black hole and the topological dilaton black hole. The thermodynamic behaviour of the solution is found to be similar to that of the Lifshitz black hole, where a phase transition may occur for sufficiently small charge in the canonical ensemble, or sufficiently small potential in the grand canonical ensemble.
\end{abstract}
%

\section{Introduction} \label{intro}

In the various applications of General Relativity, spacetimes which are non-asymptotically flat have gained increasing interest. Perhaps the most notable is the asymptotically Anti-de Sitter spacetimes which play an important role in the gauge/gravity correspondence, string theory, and quantum gravity. Extensions of these and other related ideas have subsequently led to spacetimes of other asymptotics.

For instance, attention has recently been turned to spacetimes which may serve as a gravitational dual to non-relativistic field theories \cite{Son:2008ye,Balasubramanian:2008dm}. For a more detailed review, see, e.g., \cite{Taylor:2008tg,Taylor:2015glc,Park:2014raa}. Here we shall briefly recall the essential ideas that are most relevant to this paper. Such non-relativistic field theories in $d$-dimensions require an anisotropic scaling between space and time in the form 
\begin{align}
 t\rightarrow \lambda^zt,\quad \vec{x}\rightarrow\lambda\vec{x}, \label{Lifshitz_scaling}
\end{align}
where $\lambda$ is a real constant, $t$ denotes the time coordinate and $\vec{x}$ denotes the $(d-1)$-dimensional spatial coordinates. The real constant $z$ is referred to as the \emph{Lifshitz exponent}. 


The scaling requirement leads the authors of \cite{Kachru:2008yh,Balasubramanian:2008dm} to consider gravity duals taking the form of the \emph{Lifshitz spacetime}. Concretely, a particular form of the Lifshitz spacetime in $D=(d+1)$-dimensions is given by 
\begin{align}
 \dif s^2&=-\frac{r^{2z}}{L^2}\dif t^2+\frac{L^2}{r^2}\dif r^2+r^2\dif\vec{x}^2_{(d-1)}, \label{Lifshitz_flat}
\end{align}
where $L$ is a constant curvature scale. Clearly this spacetime is invariant under the scaling 
\begin{align}
 t\rightarrow\lambda^zt,\quad r\rightarrow\frac{r}{\lambda},\quad \vec{x}\rightarrow\lambda\vec{x},
\end{align}
which satisfies Eq.~\Eqref{Lifshitz_scaling}, along with an additional coordinate $r$ representing the one extra dimension of the holographic dual. Furthermore, duals to systems at finite temperature require the presence of a black hole. Hence on the gravity side, a suitable holographic dual would be the \emph{Lifshitz black hole}, which are black holes in spacetimes which are asymptotically \Eqref{Lifshitz_flat}. 

On the gravity side, work has been done to find an appropriate action which supports the Lifshitz spacetime as a solution to its equations of motion. Such a solution require the presence of various matter fields. The approach adopted by \cite{Balasubramanian:2008dm,Balasubramanian:2009rx,Danielsson:2009gi,Mann:2009yx} is to consider gravity a coupled to one-form and two-form gauge fields. Construction of the solutions using Einstein--Proca theory was achieved in \cite{Bertoldi:2009vn,Liu:2014dva}. The Lifshitz black hole can also be constructed in Lovelock gravity \cite{Dehghani:2010kd} and higher-curvature gravity \cite{AyonBeato:2009nh,Gim:2014nba}. 
 
Of relevence to this paper is the thread where an Einstein--Maxwell--dilaton (EMd)-type action is used. In Ref.~\cite{Li:2016rcv}, such an EMd-type action with multiple $U(1)$ gauge fields and multiple dilaton fields are used to construct hyperscaling-violating solutions with Lifshitz scaling.
On the other hand, an EMd-type action with multiple $U(1)$ fields with a single dilaton was considered by Tarrio and Vandoren \cite{Tarrio:2011de} to construct a charged spherical black hole in Lifshitz spacetime, generalising Eq.~\Eqref{Lifshitz_flat}. There, the authors have found that distinct $U(1)$ fields are required to support the various curvature structures of the solution. 

To elaborate on this point, we take a look at the action used by Tarrio and Vandoren \cite{Tarrio:2011de}, for which the bulk term in $D=(d+1)$-dimensions is
\begin{align}
 I&=\frac{1}{16\pi G}\int\dif^Dx\sqrt{-g}\brac{R-2\Lambda-\half\brac{\nabla\psi}^2-\frac{1}{4}\sum_{i=1}^N\expo{-2a_i\psi}F_i^2}, \label{Lifshitz_action}
\end{align}
where $G$ is the $D$-dimensional gravitational constant and $\Lambda$ is the cosmological constant. The gauge fields $F_i$ represent a distinct $U(1)$ field for each $i$, each of them coupled to the same dilaton field $\psi$ through their respective coupling parameter $a_i$. It was known since \cite{Taylor:2008tg} that at least one $U(1)$ field (say, $F_1$) is necessary to support the Lifshitz spacetime \Eqref{Lifshitz_flat} as a solution, even in the absence of the black hole. More specifically, if $F_1$ vanishes, Eq.~\Eqref{Lifshitz_flat} will reduce to the Anti-de Sitter spacetime with $z=1$. If one wishes to include a black hole with a spherical horizon, a second $U(1)$ field $F_2$ is needed. Finally, a third gauge field $F_3$ is used to charge up the black hole itself. Subsequent inclusions of $F_4, F_5,\ldots,F_N$ will contribute to different $U(1)$ charges on the black hole. The value of $z$ fixes $F_1$, while the positive curvature of the black hole's spherical horizon fixes $F_2$. Hence, out of $N$ distinct gauge fields, only $N-2$ of them ($F_3,F_4,\ldots,F_N$) are freely parameterised (within constraints) charges carried by the black hole itself. The holographic and thermodynamic consequences of these solutions were studied in \cite{Langley:2018msz,Dayyani:2018mmm}, in addition to the original analysis performed by \cite{Tarrio:2011de}.

Solutions for EMd-type theories have also been of interest in the context of string theory and supergravity, especially those involving dimensional compactification \cite{Horava:1996ma,Lukas:1998yy}. These studies involve bulk actions of the form 
\begin{align}
 I&=\frac{1}{16\pi G}\int\dif^Dx\sqrt{-g}\brac{R-\half\brac{\nabla\psi}^2-\frac{1}{4}\expo{-2a\psi}F^2-2\Lambda_1\expo{2b_1\psi}-2\Lambda_2\expo{2b_2\psi}}. \label{EMD_action}
\end{align}
In other words, this action contains a single $U(1)$ gauge field $F$, and the scalar dilaton $\psi$ has a Liouville-type potential consisting of exponetial terms $\Vcal(\psi)=2\Lambda_1\expo{2b_1\psi}+2\Lambda_2\expo{2b_2\psi}$, where $\Lambda_1$, $\Lambda_2$, $b_1$ and $b_2$ are constants. Black hole solutions under this action have been studied, for instance, in Refs.~\cite{Bousso:1996pn,Poletti:1994ff,Chan:1995fr,Cai:1997ii,Cai:1999xg}, along with multi-black holes in a cosmological background \cite{Maki:1992tq}, as well as magnetised universes with and without a black hole \cite{Radu:2003av,Agop:2005np}. 
%

For these solutions, a similar intuition can be carried over from the Lifshitz black hole with multiple gauge fields. Though in this case the situation is more straightforward; the value of $\Lambda_1$ determines whether the horizon has positive, negative, or zero curvature. Here, we take the horizon to be the surface which separates two regions of a spacetime, in which time-like and null curves are unable to escape from one region into the other. (See, e.g., \cite{Cai:1996eg,Vanzo:1997gw,Birmingham:1998nr,Cai:1997ii}.) Therefore, the value of $\Lambda_1$ determines the curvature of this surface. Finally, $F$ contributes to the charge of the black hole. Further properties and thermodynamics of these solutions were studied by \cite{Cai:1997ii,Cai:1999xg,Agop:2005np}. More general features of spacetimes under this potential were considered in Refs.~\cite{Charmousis:2001nq,Charmousis:2009xr,Abdolrahimi:2011zg}.

In the persent paper, we wish to study the possible solution of a action that contains the ingredients of both \Eqref{Lifshitz_action} and \Eqref{EMD_action}, so that the Lifshitz and dilaton black hole solutions are contained within a single framework. Specifically, the action is an EMd-type action with $M$ Liouville potentials and $N$ gauge fields, for any number $M$ and $N$. The first hints of this possibility may come from noticing that the Lifshitz black hole and dilaton black hole solution have certain features in common with each other. (For example, that the dilaton $\psi$ in both cases is proportional to $\ln r$.) 

The solution presented in this paper is spacetime which contains the Lifshitz and dilaton black holes, where a parameter $\nu$ that interpolates between the two. The horizon of the black hole may have planar, spherical, or hyperbolic topology, depending on the relative values of the gauge fields and Liouville potentials. Particularly, in the case of the planar horizon, the black hole asymptotically approaches \Eqref{Lifshitz_flat}. Another consequence of the solution is that the Lifshitz black hole metric of \cite{Tarrio:2012xx} can be reproduced by replacing one $U(1)$ gauge field by a Liouville potential.

This paper is organised as follows. In Sec.~\ref{solution}, the equations of motion are presented and the exact solution is derived. Some physical and geometrical properties of the solution will be studied in Sec.~\ref{physical}, followed by an elementary thermodynamic analysis in Sec.~\ref{thermodynamics}. The paper concludes in Sec.~\ref{conclusion}. We will be using units where the speed of light, Planck's constant, and Boltzmann's constant equals unity, $c=\hbar=k=1$. We shall also take the Lorentzian signature for the spacetime metric to be $(-,+,\ldots,+)$.

\section{Derivation of the solution} \label{solution}

Our goal is to consider an action for which a solution that extremises this action is a spacetime which reduces to the Lifshitz black hole as well as the dilaton black hole. The former solution is supported by a EMd-type action with $N$ gauge fields and no dilaton potential \cite{Tarrio:2012xx}, as sketched in Eq.~\Eqref{Lifshitz_action}, wheras the dilaton black holes of \cite{Chan:1995fr} are supported by an EMd-type action with a single gauge field with two dilaton potentials for $\psi$, sketched in Eq.~\Eqref{EMD_action}. 

Hence, in combining the features of Eqs.~\Eqref{Lifshitz_action} and \Eqref{EMD_action}, we consider an EMd-type theory consisting of $N$ different $U(1)$ gauge fields with a dilaton potential $\Vcal(\psi)$,
\begin{align}
 I&=\frac{1}{16\pi G}\int_{\mathcal{M}}\dif^Dx\sqrt{-g}\brac{R-\half\brac{\nabla\psi}^2-\frac{1}{4}\sum_{i=1}^N\expo{-2a_i\psi}F_i^2-\Vcal(\psi)}\nonumber\\
   &\hspace{2cm}+\frac{1}{8\pi G}\int_{\partial\mathcal{M}}\dif^{D-1}x\sqrt{-\gamma}\,K. \label{action}
\end{align}
where $F_i$ are the $N$ two-form fields which arise from the exterior derivative of their respective one-form potentials, $F_i=\dif A_i$. We have also denoted $F_i^2=(F_i)_{\mu\nu}(F_i)^{\mu\nu}$. Each of these fields are coupled to the scalar field $\psi$ via their respective coupling parameters $a_i$.\footnote{For comparison with Tarrio and Vandoren's action \cite{Tarrio:2011de}, we have $\lambda_i=-2a_i$ where $\lambda_i$ is the coupling parameter used in their paper.}  As in Eq.~\Eqref{Lifshitz_action} and \Eqref{EMD_action}, $G$ denotes the $D$-dimensional gravitational constant.

The second term in the action is the Gibbons--Hawking--York boundary term \cite{York:1972sj,Gibbons:1976ue}. This term serves to fix the metric on the boundary and to ensure that the variation of the action vanishes on-shell. A time-like boundary $\partial\mathcal{M}$ is chosen so that a Hamilton--Jacobi-type analysis can be applied to the action. 
Concretely, let the metric on the boundary be $\gamma_{\mu\nu}$. 
The boundary stress tensor is obtained from the variation of action with respect to $\gamma_{\mu\nu}$.  According to the analysis by Brown and York \cite{Brown:1992br}, the time-like components of this boundary stress tensor is interpreted as the quasi-local energy  contained within the slice of $\partial\mathcal{M}$ that is orthogonal to the time-like direction. For a concrete example, consider the Schwarzschild solution in its usual coordinates $(t,r,\theta,\phi)$. For a spherical volume $r<r_b$ for some chosen $r_b$, the quasilocal energy at some fixed time is the energy contained within that spherical region.

We shall consider potentials $\Vcal(\psi)$ where it is the sum of $M$ exponential terms,
\begin{align}
 \Vcal(\psi)&=\sum_{j=1}^M2\Lambda_j\expo{2b_j\psi},
\end{align}
where $b_j$ and $\Lambda_j$ are constants. Therefore, the case of $N>0$ and $M=0$ corresponds to Eq.~\Eqref{Lifshitz_action}  under which the Lifshitz black hole was obtained \cite{Tarrio:2011de}, where at least $N=3$ is needed to have a charged Lifshitz black hole with a spherical horizon.\footnote{Note the $N=2$ case in Sec.~2.1 of \cite{Tarrio:2011de} gives a black hole with a planar horizon.} The case $N=1$ and $M=2$ corresponds to Eq.~\Eqref{EMD_action} where the dilaton black hole was found \cite{Chan:1995fr}. Therefore, we should consider at least $M=N=3$ to be the simplest non-trivial solution to go beyond the cases already contained in \cite{Tarrio:2011de} and \cite{Chan:1995fr}. 

Extremising the action gives the Einstein--Maxwell--dilaton equations
\begin{align}
 &R_{\mu\nu}=\half\nabla_\mu\psi\nabla_\nu\psi+\frac{1}{D-2}\Vcal g_{\mu\nu}+\sum_{i=1}^N\sbrac{\half\expo{-2a_i\psi}(F_i)_{\mu\lambda}{(F_i)_\nu}^\lambda-\frac{1}{4(D-2)}\expo{-2a_i\psi}F_i^2 g_{\mu\nu}},\label{Einstein_Eq}\\
 &\nabla_\mu\brac{\expo{-2a_i\psi}F_i^{\mu\nu}}=0,\label{Maxwell_Eq}\\
 &\nabla^2\psi=2\frac{\dif\Vcal}{\dif\psi}-\frac{1}{2}\sum_{i=1}^Na_i\expo{-2a_i\psi}F_i^2.\label{dilaton_Eq}
\end{align}
As briefly alluded to in the Introduction, the general form of the Lifshitz and dilaton black hole share may similar features. In particular, their metrics in both cases take the form
\begin{align}
 \dif s^2&=-f(r)\dif t^2+h(r)\dif r^2+r^2\tilde{\gamma}_{ab}\dif y^a\dif y^b, \label{metric_ansatz}
\end{align}
where $f(r)$ and $h(r)$ are functions depending only on $r$. The metric $\tilde{\gamma}_{ab}\dif y^a\dif y^b$ is a $(D-2)$-dimensional space of constant unit curvature $k=0,\pm 1$. Our present goal can then be focused into seeking a solution which also takes the form \Eqref{metric_ansatz} and find an appropriate $f(r)$ and $h(r)$ that extremises the action \Eqref{action}. We shall also take the ansatz where the gauge potentials only have `electric' radial components,
\begin{align}
 A_i&=\chi_i(r)\,\dif t,
\end{align}
where $\chi_i(r)$ are scalar functions that depend only on $r$. Similarly we shall also take the ansatz where $\psi=\psi(r)$ is a scalar function that depends only on $r$. 

In the following paragraphs we shall show the details of how the solution is derived. The reader uninterested in such details may skip ahead to the final solution shown in Eqs.~\Eqref{soln_ansatz}, \Eqref{soln_alphas}, and \Eqref{soln_Lambdas}.

Substitution of the ansatz into Eq.~\Eqref{Einstein_Eq} gives
\begin{align}
 -\frac{1}{2\sqrt{fh}}\brac{\frac{f'}{\sqrt{fh}}}'-\frac{(D-2)f'}{2rfh}&=\sum_{j=1}^M\frac{2\Lambda_j\expo{2b \psi}}{D-2}-\sum_{i=1}^N\frac{D-3}{2(D-2)}\frac{1}{fh}\expo{-2 a_i\psi}\chi_i'^2,\label{EE1a}\\
 -\frac{1}{2\sqrt{fh}}\brac{\frac{f'}{\sqrt{fh}}}'+\frac{(D-2)h'}{2rh^2}&=\sum_{j=1}^M\frac{2\Lambda_j\expo{2 b_j\psi}}{D-2}-\sum_{i=1}^N\frac{D-3}{2(D-2)}\frac{1}{fh}\expo{-2 a_i\psi}\chi_i'^2+\frac{1}{2h}\psi'^2,\label{EE2a}\\
 \frac{D-3}{r^2}\brac{k-\frac{1}{h}}+\frac{1}{2rh}\brac{\frac{h'}{h}-\frac{f'}{f}}&=\sum_{j=1}^M\frac{2\Lambda_j\expo{2b_j\psi}}{D-2}+\sum_{i=1}^N\frac{1}{2(D-2)}\frac{1}{fh}\expo{-2a_i\psi}\chi_i'^2,\label{EE3a}
\end{align}
where primes denote derivatives with respect to $r$. Furthermore, putting the ansatz into the Maxwell equation \Eqref{Maxwell_Eq} and dilaton equation \Eqref{dilaton_Eq} give
\begin{align}
 \brac{\frac{r^{D-2}}{\sqrt{fh}}\expo{-2 a_i\psi}\chi_i'}'&=0,\quad i=1,\ldots, N,\label{Ma}\\
 \frac{1}{\sqrt{fh}r^{D-2}}\brac{\sqrt{\frac{f}{h}}r^{D-2}\psi'}'&=\sum_{j=1}^M  4b_j\Lambda_j\expo{2 b_j\psi}+\sum_{i=1}^N\frac{ a_i}{fh}\expo{-2 a_i\psi}\chi_i'^2.\label{Da}
\end{align}
Taking the difference between \Eqref{EE1a} and \Eqref{EE2a} leads to
\begin{align}
 \frac{D-2}{r}\brac{\frac{h'}{h}+\frac{f'}{f}}&=\psi'^2.\label{varphi_eqn}
\end{align}
To make progress towards obtaining a solution, we shall follow \cite{Chan:1994qa,Chan:1995fr,Cai:1997ii} and consider a specific ansatz for $h$ as
\begin{align}
 h=\frac{r^{2n}}{f}, \label{h}
\end{align}
for some constant $n$. While this does not rule out the possibility that there exist other of solution for other forms of $h$, the ansatz taken here will nevertheless lead to a solution for $f$ that interpolates between the Lifshitz and dilaton black hole cases. With this ansatz for \Eqref{h}, Eq.~\Eqref{varphi_eqn} becomes $\psi'^2=2(D-2)/r^2$ and can be solved to give\footnote{Following \cite{Cai:1997ii,Chan:1994qa,Tarrio:2011de}, we shall choose the negative sign upon taking the square root.  As we will see below, solving the equations of motion for this choice of sign will lead to a positive sign for the coupling parameter $a_i$ when the solution to reduced to that of \cite{Chan:1994qa} and \cite{Tarrio:2011de}. Furthermore this will be consistent if the terms $\expo{-2a_i\psi}F_i^2$ of the action is a result of a dimensional compactification \cite{Poletti:1994ff,Horava:1996ma,Lukas:1998yy}, in which case $a_i$ is a positive number that depends on the number of reduced dimensions.} 
\begin{align}
 \psi=-\delta\ln r+\psi_0,\quad \delta=\sqrt{2(D-2)n},\label{dilaton}
\end{align}
where $\psi_0$ is an integration constant. It is worth noting that besides the context of dilaton and Lifshitz black hole solutions, scalar fields of the form \Eqref{dilaton} are solutions to asymptotically-AdS spacetimes in various models \cite{Anabalon:2013eaa,Acena:2013jya,Anabalon:2017yhv}. 

The Maxwell equations for each $U(1)$ gauge field \Eqref{Ma} can then be immediately integrated once to give 
\begin{align}
 \chi_i'&=\lambda_i\expo{ a_i\psi_0}r^{-(D-2-n+2 a_i\delta)}, \label{potential}
\end{align}
where each $\lambda_i$ are the integration constants parametrising the strength of their respective $U(1)$ charges.

Substituting Eqs.~\Eqref{h}, \Eqref{dilaton}, and \Eqref{potential} into Eqs.~\Eqref{EE1a}, \Eqref{EE3a}, and \Eqref{Da} leads to 
\begin{subequations}\label{Romans}
\begin{align}
 &-\frac{1}{2r^n}\brac{\frac{f'}{r^n}}'-\frac{(D-2)f'}{2r^{2n+1}}=\frac{2}{D-2}\sum_{j=1}^M V_jr^{-2 b_j\delta}-\frac{D-3}{2(D-2)}\sum_{i=1}^N \lambda_i^2r^{-2 a_i\delta-2(D-2)},\label{I}\\
 &(D-3)kr^{-2}-(D-3-n)r^{-2-2n}f-r^{-1-2n}f'=\frac{2}{D-2}\sum_{j=1}^M V_jr^{-2 b_j\delta}\nonumber\\
 &\hspace{9cm}+\frac{1}{2(D-2)}\sum_{i=1}^N \lambda_i^2r^{-2 a_i\delta-2(D-2)},\label{II}\\
 &-\delta r^{-(D-2+n)}\brac{r^{D-3-n}f}'=4\sum_{j=1}^M b_jV_jr^{-2 b_j\delta}+\sum_{i=1}^N  a_i\lambda_i^2r^{-2 a_i\delta-2(D-2)},\label{III}
\end{align}
\end{subequations}
where we have denoted
\begin{align}
 V_j=\Lambda_j\expo{2 b_j\psi_0}.
\end{align}

What remains is to determine $f$. We shall motivate our ansatz by noting that, for example, the Reissner--Nordstr\"{o}m-AdS solution has the form $f=k-\frac{\mu}{r^{D-3}}+\frac{q^2}{r^{2(D-2)}}+\frac{r^2}{\ell^2}$, particularly that the term $\frac{q^2}{r^{2(D-2)}}$ is associated with the $U(1)$ gauge field and the term $\frac{r^2}{\ell^2}$ is associated with the cosmological constant. Either of these terms vanish if their respective gauge field or cosmological constant is switched off. A negative cosmological constant allows for $k=0$ or $k=-1$. A similar situation occurs for the charged Lifshitz black hole with $N$ $U(1)$ gauge fields \cite{Tarrio:2011de}, where their function $f$ has a separate term for each of the $N$ gauge fields. One of the gauge fields is required in order to have a spherical black hole. In the dilaton black holes of \cite{Chan:1995fr,Cai:1999xg}, there is a separate term for each Liouville potential. 

We then carry these considerations into our present context, which has $N$ gauge fields and $M$ Liouville potentials. To this end, first we rewrite the sums appearing in the right-hand sides of Eq.~\Eqref{Romans} explicitly as three parts, for instance,
\begin{subequations}\label{sumsplit}
\begin{align}
 \sum_{i=1}^N \lambda_i^2 r^{-2 a_i\delta-2(D-2)}&=\lambda_N^2 r^{-2 a_N\delta-2(D-2)}+\sum_{l=1}^p \lambda_l^2 r^{-2 a_l\delta-2(D-2)}+\sum_{i=p+1}^{N-1}\lambda_i^2,
 r^{-2 a_i\delta-2(D-2)},\\
 \sum_{j=1}^M\frac{2V_j}{D-2}r^{-2 b_j\delta}&=\frac{2V_M}{D-2}r^{-2 b_M\delta}+\sum_{l=1}^{p}\frac{2V_l}{D-2}r^{-2 b_l\delta}+\sum_{j=p+1}^{M-1}\frac{2V_j}{D-2}r^{-2 b_j\delta}.
\end{align}
\end{subequations}
Next, we take our ansatz for $f$ to have $M+N-p$ terms, 
\begin{align}
 f=Br^{\gamma}-\mu r^{c_0}+\sum_{l=1}^p\frac{r^{\theta_l}}{L_l^2}+\sum_{i=p+1}^{N-1}q_i^2r^{\omega_i}+\sum_{j=p+1}^{M-1}\frac{r^{\sigma_j}}{\ell_j^2},\label{f_ansatz}
\end{align}
where the constant coefficients $\frac{1}{L_l^2}$, $q_i^2$, $\frac{1}{\ell_j^2}$, $B$, and constant exponents $c_0$, $\theta_l$, $\omega_i$, $\sigma_j$, and $\gamma$ are to be determined. The reason why we consider $M+N-p$ terms is to include the possibility of overlap; namely that $p$ of the terms receive simultaneous contributions from a gauge field and Liouville potential, for some $p$. We will also let the term $Br^\gamma$ be sourced by the $N$-th gauge field and $M$-th Liouville potential. The term $-\mu r^{c_0}$ is a Schwarzschild-like term and is not sourced by any gauge field or Liouville potential.   

In the following, we shall fix the indices $l$, $i$, and $j$ to run according to the range implied in Eq.~\Eqref{sumsplit} and \Eqref{f_ansatz}.

Substitution of Eq.~\Eqref{f_ansatz} into \Eqref{I}, along with splitting the sums according to \Eqref{sumsplit} gives 
\begin{align}
 &-\half(D-3-n+\gamma)\gamma Br^{\gamma-2-2n}-\half\sum_{l=1}^p(D-3-n+\theta_l)\frac{\theta_l}{L_l^2}r^{\theta_l-2-2n}\nonumber\\
 &-\half\sum_{i=p+1}^{N-1}(D-3-n+\omega_i)\omega_iq_i^2r^{\omega_i-2-2n}-\half\sum_{j=p+1}^{M-1}(D-3-n+\sigma_j)\frac{\sigma_j}{\ell_j^2}r^{\sigma_j-2-2n} \nonumber\\
 &-\half(D-3-n+c_0)c_0\mu r^{c_0}\nonumber\\
 &\hspace{2cm}=\,\frac{2V_M}{D-2}r^{-2 b_M\delta}-\frac{(D-3) \lambda_M^2}{2(D-2)}r^{- a_M\delta-2(D-2)}\nonumber\\
 &\hspace{2.5cm}+\sum_{l=1}^p\sbrac{\frac{2V_l}{D-2}r^{-2 b_l\delta}-\frac{(D-3) \lambda_l^2}{2(D-2)}r^{-2 a_l\delta-2(D-2)}}\nonumber\\
 &\hspace{2.5cm}-\sum_{i=p+1}^{N-1}\frac{(D-3) \lambda_i^2}{2(D-2)}r^{-2 a_i\delta-2(D-2)}+\sum_{j=p+1}^{M-1}\frac{2V_j}{D-2}r^{-2 b_j\delta}.\label{Ia}
\end{align}
Substitution of Eq.~\Eqref{f_ansatz} into \Eqref{II} and splitting the sums gives
\begin{align}
 &(D-3)kr^{-2}-(D-3-n+\gamma)Br^{\gamma-2-2n}\nonumber\\
 &-\sum_{l=1}^p(D-3-n+\theta_l)L_l^{-2}r^{\theta_l-2-2n}-\sum_{i=p+1}^{N-1}(D-3-n+\omega_i)q_i^2r^{\omega_i-2-2n}\nonumber\\
 &-\sum_{j=p+1}^{M-1}(D-3-n+\sigma_j)\ell_j^{-2}r^{\sigma_j-2-2n}\nonumber\\
 &-(D-3-n-c_0)\mu r^{c_0-2-2n}\nonumber\\
 &\hspace{2cm}=\,\frac{2 V_M}{D-2}r^{-2 b_M\delta}+\frac{\lambda_N^2}{2(D-2)}r^{-2 a_N-2(D-2)}\nonumber\\
 &\hspace{2.5cm}+\sum_{l=1}^p\sbrac{\frac{2 V_l}{D-2}r^{-2 b_l\delta}+\frac{\lambda_l^2}{2(D-2)}r^{-2 a_l\delta-2(D-2)}}\nonumber\\
 &\hspace{2.5cm}+\sum_{i=p+1}^{N-1}\frac{ \lambda_i^2}{2(D-2)}r^{-2 a_i\delta-2(D-2)}+\sum_{j=p+1}^{M-1}\frac{2 V_j}{D-2}r^{-2 b_j\delta}. \label{IIa}
\end{align}
Substitution of Eq.~\Eqref{f_ansatz} into \Eqref{III} and splitting the sums gives 
\begin{align}
 &-\delta(D-3-n+\gamma)Br^{\gamma-2n-2}\nonumber\\
 &-\sum_{l=1}^p\delta(D-3-n+\theta_l)L_l^{-2}r^{\theta_l-2n-2}-\sum_{i=p+1}^{N-1}\delta(D-3-n+\omega_i)q_i^2r^{\omega_i-2n-2}\nonumber\\
 &-\sum_{j=p+1}^{M-1}\delta(D-3-n+\sigma_j)\ell_j^{-2}r^{\sigma_j-2n-2}-\delta(D-3-n+c_0)\mu r^{c_0-2n-2}\nonumber\\
 &\hspace{2cm}=\,4 b_M V_Mr^{-2 b_M\delta}+ a_N \lambda_N^2r^{-2 a_N\delta-2(D-2)}\nonumber\\
 &\hspace{2.5cm}+\sum_{l=1}^p\sbrac{4 b_lV_l r^{-2 b_j\delta}+ a_l \lambda_i^2r^{-2 a_l\delta-2(D-2)}}\nonumber\\
 &\hspace{2.5cm}+\sum_{i=p+1}^{N-1}  a_i\lambda_i^2r^{-2 a_i\delta-2(D-2)}+\sum_{j=p+1}^{M-1}4 b_j V_j r^{-2 b_j\delta}. \label{IIIa}
\end{align}

Firstly, if $c_0=-(D-3-n)$, then the terms involving $\mu$ vanishes from the equations and in $f$, it appears as 
\begin{align}
 -\mu r^{-(D-3-n)},
\end{align}
which is the Schwarzschild-like term and is associated with the presence of the black hole.

The rest of the equations are solved via term-by-term comparisons.  The remaining terms in the left-hand sides of these equations involve the coefficients $k$, $B$, $\frac{1}{L_l^2}$, $q_i^2$, and $\frac{1}{\ell_j^2}$, while the right-hand sides of the equations involve $\lambda_1,\ldots,\lambda_p,\lambda_{p+1},\ldots,\lambda_{N-1},\lambda_N$, and $V_1,\ldots,V_p,V_{p+1},\ldots,V_{M-1},V_M$. We match the terms on both sides of each equations by the following scheme:
\begin{align}
 \brac{\mbox{terms involving }L_l^{-2}}&\leftrightarrow\brac{\mbox{terms involving } \lambda_l,V_l},\quad l=1,\ldots,p,\nonumber\\
 \brac{\mbox{terms involving }q_i}&\leftrightarrow\brac{\mbox{terms involving }\lambda_i},\quad i=p+1,\ldots,N-1\nonumber\\
 \brac{\mbox{terms involving }\ell_j^{-2}}&\leftrightarrow\brac{\mbox{terms involving } V_j},\quad j=p+1,\ldots,M-1,\nonumber\\
 \brac{\mbox{terms involving }k,B}&\leftrightarrow\brac{\mbox{terms involving } \lambda_N,V_M}.
\end{align}
For example, matching terms involving $q_i$ with terms containing $\lambda_i$ in Eq.~\Eqref{Ia}, \Eqref{IIa}, and \Eqref{IIIa} respectively gives 
\begin{align}
 -\half(D-3-n+\omega_i)\omega_iq_i^2r^{\omega_i-2-2n}&=-\frac{(D-3)\lambda_i^2}{2(D-2)}r^{- a_i\delta-2(D-2)},\\
 -(D-3-n+\omega_i)q_i^2r^{\omega_i-2-2n}&=\frac{\lambda_i^2}{2(D-2)}r^{-2 a_i\delta-2(D-2)},\\
 -\delta\brac{D-3-n+\omega_i}q_i^2r^{\omega_i-2-2n}&= a_i \lambda^2_i r^{-2 a_i\delta-2(D-2)}.
\end{align}
Demading that the coefficients and exponents of $r$ on both sides of each equations to be equal gives 
\begin{align}
 \omega_i&=-2(D-3),\quad  a_i\delta=n,\nonumber\\
 \lambda_i&=\sqrt{2(D-2)(D-3+n)}q_i.
\end{align}

Similarly, matching the terms involving $\frac{1}{L_l^2}$ with the terms containing $\lambda_l$ and $V_l$ gives three algebraic equations. Matching the coefficients and exponents of $r$, we find
\begin{align}
 \theta_l&=2+2n-2\nu_l,\quad a_l\delta=\nu_l-(D-2),\nonumber\\
 \lambda_l^2&=2(D-1+n-2\nu_l)(n-\nu_l)\frac{1}{L_l^2},\nonumber\\
 2V_l&=2\Lambda_l\expo{2b_l\psi_0}=-(D-1+n-2\nu_l)(D-2+n-\nu_l)\frac{1}{L_l^2},
\end{align}
where we have introduced the parametrisation $ b_l\delta=\nu_l$. Next, matching the terms containing $\frac{1}{\ell_j^2}$ with terms containing $V_j$ gives 
\begin{align}
 \sigma_j&=2,\quad  b_j\delta=n,\nonumber\\
 2V_j&=2\Lambda_j\expo{2 b_j\psi_0}=-(D-2)(D-1-n)\frac{1}{\ell_j^2},
\end{align}
whereas matching the terms involving $B$ with terms containing $\lambda_N$ and $V_M$ results in equations which are solved by  
\begin{align}
 \gamma&=2n,\quad b_M\delta=1,\quad a_N\delta=-(D-3),\nonumber\\
B&=\frac{(D-3)^2k-2\Lambda_M\expo{2 b_M\psi_0}}{(D-3+n)^2}=\frac{(D-3)k-(D-3+n)\rho^2}{(1-n)(D-3+n)},
\end{align}
where $\lambda_N=\sqrt{2(D-3+n)}\rho$.

The gauge potentials are obtained upon integrating the Maxwell equations again, which gives
\begin{subequations}
\begin{align}
 \chi_l&=\Phi_l+\expo{\alpha_l\psi_0}\sqrt{\frac{2(n-\nu_l)}{D-1+n-2\nu_l}}\frac{r^{D-1+n-2\nu_l}}{L_l},\\
 \chi_i&=\Phi_i-\expo{\alpha_i\psi_0}\sqrt{\frac{2(D-2)}{D-3+n}}q_ir^{-(D-3+n)},\\
 \chi_N&=\Phi_N+\expo{\alpha_N\psi_0}\sqrt{\frac{2}{D-3+n}}\rho r^{D-3+n}.
\end{align}
\end{subequations}
where, $\Phi_l$, $\Phi_i$, and $\Phi_N$ above are the integration constants.

We collect and summarise our results: Reiterating for the benefit of the reader who has skipped the details of the derivation, the indices $l$, $i$, and $j$, appearing in the equations below are taken to run as follows
\begin{align}
 l\in\{1,\ldots,p\},\quad i\in\{p+1,\ldots,N-1\},\quad j\in\{p+1,\ldots,M-1\}.
\end{align}
The metric, gauge potentials and dilaton field which solves the equations of motion \Eqref{Einstein_Eq}, \Eqref{Maxwell_Eq}, and \Eqref{dilaton_Eq} are 
\begin{subequations}\label{soln_ansatz}
\begin{align}
 \dif s^2&=-f\dif t^2+h\dif r^2+r^2\dif\Omega^2, \\
 f&=B r^{2n}-\frac{\mu}{r^{D-3-n}}+\sum_{l=1}^p\frac{r^{2+2n-2\nu_l}}{L_l^2}+\sum_{i=p+1}^{N-1}\frac{q_i^2}{r^{2(D-3)}}+\sum_{j=p+1}^{M-1}\frac{r^2}{\ell_j^2}, \label{f_soln}\\
 h&=\frac{r^{2n}}{f},\label{g_soln}\\
 A_l&=\brac{\Phi_l+\expo{a_l\psi_0}\sqrt{\frac{2(n-\nu_l)}{D-1+n-2\nu_l}}\frac{r^{D-1+n-2\nu_l}}{L_l}}\,\dif t,\\
 A_i&=\brac{\Phi_i-\expo{-a_i\psi_0}\sqrt{\frac{2(D-2)}{D-3+n}}q_ir^{-(D-3+n)}}\,\dif t,\\
 A_N&=\brac{\Phi_N+\expo{a_N\psi_0}\sqrt{\frac{2}{D-3+n}}\rho r^{D-3+n}}\,\dif t,\\
 \psi&=-\sqrt{2(D-2)n}\ln r+\psi_0,
\end{align}
\end{subequations}
provided that the Liouville potential strengths are related to $B$, $\frac{1}{L_l^2}$, $\frac{1}{\ell_j^2}$ by
\begin{subequations}\label{soln_Lambdas}
\begin{align}
 2\Lambda_l\expo{2 b_l\psi_0}&=-\frac{(D-1+n-2\nu_l)(D-2+n-\nu_l)}{L_l^2},\label{Lambda_l}\\
 2\Lambda_j\expo{2 b_j\psi_0}&=-\frac{(D-2)(D-1-n)}{\ell_j^2},\label{Lambda_j}\\
 B&=\frac{(D-3)^2k}{(D-3+n)^2}+\frac{2\Lambda_M\expo{2 b_M\psi_0}}{(D-3+n)^2}, \label{rhodef}
\end{align}
\end{subequations}
and that the coupling constants are related to the parameters $\nu_l$ and $n$ by
\begin{align}
 a_l&=\frac{\nu_l-(D-2)}{\sqrt{2(D-2)n}},\quad
 a_i=\sqrt{\frac{n}{2(D-2)}},\quad
 a_N=-\sqrt{\frac{D-3}{2(D-2)n}}, \nonumber\\
 b_l&=\frac{\nu_l}{\sqrt{2(D-2)n}},\quad b_j=\sqrt{\frac{n}{2(D-2)}},\quad b_M=\frac{1}{\sqrt{2(D-2)n}}.\label{soln_alphas}
\end{align}

We recover known solutions by the following choices of the parameters:
\begin{itemize}
 \item \textbf{Lifshitz spherical/planar black hole}: The Lifshitz black hole is recovered by setting $\nu_l=0$ and $\Lambda_j=0$. The latter condition is tantamount to having $\ell_j^2\rightarrow\infty$ and all the terms proportional to $r^{2}$ in $f$ will vanish.  
 Then, with $\nu_l=0$, the $p$ gauge fields are redundant since they all give rise to the same term proportional to $r^{2+2n}$, so we might well consider $p=1$. If we further let $z=n+1$, the function $f$ can be rewritten as 
 \begin{align}
  f=r^{2z}\sbrac{\frac{(D-3)^2k}{(D-3+n)^2r^2}+\frac{1}{L^2}-\frac{\mu}{r^{D-2-z}}+\sum_{i=2}^{N-1}q_i^2r^{-2(D-3+z)}}.
 \end{align}
 We see that the term in the square brackets is precisely the function denoted as $b_k$ in \cite{Tarrio:2011de} and that $z$ is the familiar Lifshitz exponent. Along with rest of the solution, we have recovered the Lifshitz spherical/planar black hole derived by \cite{Tarrio:2011de}.
 
 \item \textbf{The dilaton black hole}: If we set $\nu_l=n$, then the gauge fields $\chi_l$ where $l=1,\ldots,p$ are switched off. Then, Eq.~\Eqref{Lambda_l} becomes the same as Eq.~\Eqref{Lambda_j}, $b_l=b_j=\sqrt{\frac{n}{2(D-2)}}$. Then $M-1$ terms in the dilaton potential become identical and hence redundant, because 
 \begin{align}
  \Vcal&=\sum_{l=1}^p2\Lambda_l\expo{2b_l\psi}+\sum_{i=p+1}^{M-1}2\Lambda_j\expo{2b_j\psi}+2\Lambda_M\expo{2b_M\psi}\nonumber\\
    &=2\brac{\sum_{j=1}^p\Lambda_l+\sum_{i=p+1}^{M-1}\Lambda_i}\expo{\sqrt{\frac{2n}{D-2}}\psi}+2\Lambda_M\expo{2 b_M\psi}.
 \end{align}
 We can then rename $\sum_{j=1}^p\Lambda_l+\sum_{i=p+1}^{M-1}\Lambda_i=\Lambda_1$, and $M=2$ suffices. We then have 
 \begin{align}
  2\Lambda_1\expo{2 b_1\psi_0}=-(D-2)(D-1-n)\ell^{-2}, \quad b_1=\sqrt{\frac{n}{2(D-2)}}.
 \end{align}
 Further introduding the transformation 
 \begin{align}
  r=\varrho^{\frac{1}{1+n}},\quad t=(1+n)\hat{t},
 \end{align}
 the solution becomes 
 \begin{align}
  \dif s^2&=-U\dif\hat{t}^2+U^{-1}\dif\varrho^2+\varrho^{\frac{1}{1+n}}\dif\Omega_{k,(D-2)}^2,\\
  U&=(1+n)^2\sbrac{B\varrho^{\frac{2n}{1+n}}-\mu\varrho^{\frac{n-(D-3)}{1+n}}+\sum_{i=1}^{N-1}q_i^2\varrho^{-\frac{2(D-3)}{1+n}}+\frac{\varrho^{\frac{2}{1+n}}}{\ell^2}}.
 \end{align}
 At the moment, the solution still contains $N-1$ non-zero $U(1)$ charges. If we set all but one of them to zero, we recover the charged dilaton black hole with two exponential potentials originally obtained in Ref.~\cite{Chan:1995fr}.

\end{itemize}

The intuition we have learned in solving the equations of motion with this ansatz is that all but two terms of $f$ in \Eqref{f_soln} require a source which is either a Liouville potential, gauge field, or both. Explicitly, the terms $q_i^2/r^{2(D-3)}$ are supported by $U(1)$ gauge fields, $r^2/\ell_j^2$ are supported by Liouville potentials, and $r^{2+2n-2\nu_l}/L_l^2$ are simultaneously supported by gauge fields and Liouville potentials. Turning off any of the fields or potentials appropriately will set its corresponding terms to zero. 

The first two terms of $f$ are $Br^{2n}-\mu/r^{D-3-n}$, clearly this reduces to the usual Schwarzschild--Tangherlini form $1-\mu/r^{D-3}$ when all the matter fields are turned off. The second term here is related to the black hole mass and is not affected by the presence of the matter fields, though the presence of the $M$-th potential and $N$-th gauge field modifies the particular value of $B$ in accordance to Eq.~\Eqref{rhodef}.

\section{Physical and geometrical properties} \label{physical}

For concreteness in the rest of the paper, we shall henceforth consider the simplest charged non-trivial case $M=N=3$. Loosely speaking, this is the least number of fields which has a charged solution with a spherical horizon and non-flat/non-AdS asymptotics which contain the afore-mentioned Lifshitz black hole and dilaton black hole.

Explicitly, the solution for $M=N=3$ is given by
\begin{subequations}\label{M3N3}
\begin{align}
 \dif s^2&=-f\dif t^2+r^{2n}f^{-1}\dif r^2+r^2\tilde{\gamma}_{ab}\dif y^a\dif y^b,\label{metric}\\
 f&=B r^{2n}-\frac{\mu}{r^{D-3-n}}+\frac{r^{2+2n-2\nu}}{L^2}+\frac{q^2}{r^{2(D-3)}}+\frac{r^2}{\ell^2},\\
 \psi&=-\sqrt{2(D-2)n}\ln r+\psi_0,\label{f33_def}\\
 A_1&=\brac{\Phi_1+\expo{ a_1\psi_0}\sqrt{\frac{2(n-\nu)}{D-1+n-2\nu}}\frac{r^{D-1+n-2\nu}}{L}}\dif t,\label{A1}\\
 A_2&=\brac{\Phi_2-\expo{a_2\psi_0}\sqrt{\frac{2(D-2)}{D-3+n}}qr^{-(D-3+n)}}\dif t,\label{A2}\\
 A_3&=\brac{\Phi_3+\expo{a_3\psi_0}\sqrt{\frac{2}{{D-3+n}}}\rho r^{D-3+n}}\dif t,\label{A3}\\
 2\Lambda_1\expo{2 b_1\psi_0}&=-(D-1+n-2\nu)(D-2+n-\nu)L^{-2},\label{Lambda1}\\
 2\Lambda_2\expo{2 b_2\psi_0}&=-(D-2)(D-1-n)\ell^{-2},\label{Lambda2}\\
 B&=\frac{(D-3)^2k-2\Lambda_3\expo{2 b_3\psi_0}}{(D-3+n)^2}=\frac{(D-3)k-(D-3+n)\rho^2}{(1-n)(D-3+n)},\label{B}
\end{align}
\end{subequations}
with 
\begin{align}
 a_1&=\frac{\nu-(D-2)}{\sqrt{2(D-2)n}},\quad a_2=\sqrt{\frac{n}{2(D-2)}},\quad a_3=-\frac{D-3}{\sqrt{2(D-2)n}},\nonumber\\
 b_1&=\frac{\nu}{\sqrt{2(D-2)n}},\quad b_2=\sqrt{\frac{n}{2(D-2)}},\quad b_3=\frac{1}{\sqrt{2(D-2)n}}.
\end{align}


First, let us account for the independent parameters of this solution. The first two of these are straightforwardly $L^2$ and $\ell^2$, which fixes $\Lambda_1$ and $\Lambda_2$ via Eq.~\Eqref{Lambda1} and \Eqref{Lambda2}. If we wish to consider the cases where $\Lambda_1$ and/or $\Lambda_2$ are positive, we replace $L^2\rightarrow-L^2$ and/or $\ell^2\rightarrow-\ell^2$. Next, we have $n$ which controls the strength of $\psi$. The parameter $\nu$ determines the strength of the gauge potential $A_1$ together with $n$. The parameter $q$ determines the strength of $A_2$, and is this particular $U(1)$ field that behaves like a Reissner--Nordstr\"{o}m charge in terms of the spacetime and its thermodynamics. 

The quantities $B$, $\Lambda_3$, and $\rho$ are intertwined in the two equalities in \Eqref{B}. Hence there should be one free parameter among them. (Since $n$ has already been accounted for.) Solving for $\Lambda_3$, or $\rho$, we find 
\begin{align}
 2\Lambda_3\expo{2 b_3\psi_0}&=\frac{(D-3+n)^2\rho^2-n(D-2)(D-3)k}{1-n},\label{Lambda3_eqn}\\
 \rho^2&=\frac{(1-n)2\Lambda_3\expo{2 b_3\psi_0}+n(D-2)(D-3)k}{(D-3+n)^2}. \label{rho_eqn}
\end{align}
Without loss of generality, let us choose $\Lambda_3$ as the independent parameter. Subsequently $\rho$ will be determined from Eq.~\Eqref{rho_eqn}. With these two quantities will then fix $B$ via Eq.~\Eqref{B}. Finally, $\mu$ is not related to any of the matter fields and is obviously the Schwarzschild-like term responsible for the mass of the black hole, as we will see in further detail below. 

To identify the event horizon, we consider $r=\mathrm{constant}$ hypersurfaces. The vector normal to the hypersurface is $\partial_\mu r$, and its norm is $\partial_\mu r\partial^\mu r=g^{rr}=\frac{1}{r^{2n}}f$. We see that $r=r_+$, where $f(r_+)=0$
is a null hypersurface where time-like and null curves in $r<r_+$ cannot escape to $r>r_+$. This can be seen by transforming to Eddington--Finklestein-like coordinates and observing that light cones at $r<r_+$ are tilted inwards, indicating that future-directed null geodesics goes in the direction of decreasing $r$.  
A calculation of the curvature invariants (see Eqs.~\Eqref{RicciScalar} and \Eqref{KretschmannScalar}) show that the spacetime is regular at the horizon. For the purposes of analysis, it will be convenient to express $\mu$ in terms of $r_+$ via
\begin{align}
 \mu=\frac{r_+^{D-1+n-2\nu}}{L^2}+\frac{r_+^{D-1-n}}{\ell^2}+Br_+^{D-3+n}+q^2r_+^{-(D-3+n)},\label{mu_eqn}
\end{align}
and we shall take $r_+$ as one of the parameters of the solution, where $\mu$ can be determined from $r_+$ using Eq.~\Eqref{mu_eqn}.

Having the parameters accounted for, we shall regard our solution \Eqref{M3N3} as being parametrised by the following seven quantities: 
\begin{align}
 \brac{r_+,\,q,\,L^2,\,\ell^2,\,\Lambda_3,\,n,\,\nu},
\end{align}
where Eq.~\Eqref{A1} indicates that $\nu$ must take the range $\nu\leq n$. Also, Eq.~\Eqref{Lambda3_eqn} forbids $n=1$, unless $\Lambda_3=0$. In the ranges where all the parameters are well-defined, the function $f$ typically has three roots, one of which is our parameter $r_+$ as intended. The other two roots are denoted $r_-$ and $r_{\mathrm{c}}$, where
\begin{align}
 r_-\leq r_+< r_{\mathrm{c}},
\end{align}
where $r_-$ is non-zero if $q$ is non-zero and $r_{\mathrm{c}}$ is finite if $\Lambda_1$ and/or $\Lambda_2$ are positive. The spacetime will be static with Lorentzian signature $(-,+,\ldots,+)$ in the range $r_+<r< r_{\mathrm{c}}$ and $r<r_-$. As we will show below, the latter region contains a curvature singularity at $r=0$, and hence we shall mainly be interested in physical quantities measurable by observers at $r_+<r<r_{\mathrm{c}}$, where the curvature singularity is hidden by the horizon $r_+$. This range is understood to include the case $r_{\mathrm{c}}=\infty$ for negative $\Lambda_1$ and $\Lambda_2$.

For the various physical and thermodynamic quantities below, we first evaluate quantities in the spacetime region 
\begin{align}
 r_+<r<r_b,
\end{align}
The boundary $r=r_b$ shall be denoted as $\partial\mathcal{M}$ with unit normal
\begin{align}
 n^\mu=r_b^{-n}\sqrt{f(r_b)}\delta^\mu_r. \label{n_normal}
\end{align}
The spacetime carries a time-like Kiling vector $\xi^\mu\propto\delta^\mu_t$. It can be explicitly checked that $\xi^\mu$ solves the Killing equation if the proportionality factor is a constant. We shall choose the proportionality constant such that $\xi_\mu\xi^\mu=-1$ at the boundary. Therefore,
\begin{align}
 \xi^\mu=\frac{1}{\sqrt{f(r_b)}}\delta^\mu_t. \label{Killing_vec}
\end{align}


To calculate the mass, we adopt the boundary stress tensor procedure of \cite{Myers:1999psa} (see also \cite{Brown:1992br}), where we consider the $r=r_b$ unit normal given by Eq.~\Eqref{n_normal}. The strategy is as follows: The procedure is to calculate the boundary stress tensor which is associated with the quasi-local energy contained in the volume inside the boundary where $r_b$ at this stage is large but finite. Next one has to perform an appropriate subtraction to remove the contribution of the `background' spacetime. (Here, we shall take the `background' as the spacetime corresponding to $\mu=0$ in the metric.) The resulting stress tensor can then be regarded as coming purely from the black hole's contribution. Furthermore, upon subtraction there will no longer be terms that grow with $r_b$. Then the limit $r_b\rightarrow\infty$ then can be taken safely. Note that the scalar invariant $\xi_\mu\xi^\mu=-1$ remains finite at any $r_b$, by construction. 

The induced metric at the boundary is $\gamma_{\mu\nu}=g_{\mu\nu}-n_\mu n_\nu$. The boundary stress tensor is the variation of the action with respect to $\gamma_{\mu\nu}$ \cite{Brown:1992br}:
\begin{align}
 T_{\mu\nu}&=\frac{2}{\sqrt{|\gamma|}}\frac{\delta I}{\delta \gamma_{\mu\nu}}=-\frac{1}{8\pi G}\brac{K_{\mu\nu}-K \gamma_{\mu\nu}},
\end{align}
evaluated at some $r=r_b$. In general, the energy calculated from this stress tensor will grow with $r_b$, as it accounts for the energy of the entire spacetime. Hence we shall subtract the contribution from an appropriately chosen background and the remaining quantity is independent of $r_b$ and is to be regarded as the contribution from the black hole alone. 

To this end, we shall take our background $g_{\mu\nu}^0$ as the $\mu=0$ case of the metric. The time coordinate of the background is scaled appropriately so that $g_{\tau\tau}^0=g_{tt}$ at the boundary \cite{Myers:1999psa}. We then repeat the procedure of the above paragraph for this spacetime to obtain its corresponding stress tensor $\leftidx{^0}T_{\mu\nu}$. Indeed, the choice of background is not unique and is made to agree when the parameters coincide with cases known in earlier literature, especially Refs.~\cite{Tarrio:2011de} and \cite{Cai:1999xg}. Our background-subtracted stress tensor is obtained by 
\begin{align}
 \hat{T}_{\mu\nu}&=T_{\mu\nu}-\leftidx{^0}T_{\mu\nu}.
\end{align}
As explained in \cite{Brown:1992br}, if the Killing vector is normalised according to \Eqref{Killing_vec}, the conserved charge associated with the time-like Killing vector $\xi^\mu$ agrees with the quasi-local energy of the spacetime. Upon taking $r_b\rightarrow\infty$, the resulting constant quantity is interpreted as the `mass' of the black hole and is given explicitly by
\begin{align}
 M=\oint_{\partial\mathcal{M}}\dif^{D-2}x\sqrt{\gamma}\,\xi^\mu\xi^\nu\hat{T}_{\mu\nu}=\frac{\Omega}{16\pi G}(D-2)\mu,\label{M_def}
\end{align}
where $\Omega=\int\dif^{D-2}\sigma\sqrt{\tilde{\gamma}}$ is the volume of the part of the spacetime described by $\tilde{\gamma}_{ab}$. It is worth emphasising that all intermediate calculations are performed at finite $r_b$, and the final expressian for the scalar $M$ given in \Eqref{M_def} is ultimately independent of $r_b$ and therefore the limit $r_b\rightarrow\infty$ can be taken safely. In the topological dilaton black hole case with $\nu=n$, this agrees with the mass obtained by the counter-term method by Cai and Ohta \cite{Cai:1999xg}, and also agrees in the Lifshitz case ($\nu=0$) with the calculation by Tarrio and Vandoren \cite{Tarrio:2011de} wherein the Komar integral method was used.

In the canonical ensemble which we will consider in the thermodynamic analysis, the charge of the spacetime is fixed. In that case a more appropriate background would be the extremal spacetime $g_{\mu\nu}^e$, which is the case where $r_-$ coincides with $r_+$. Let $\mu_e$ be the mass parameter such that $r_-=r_+$. Then the mass of the black hole measured against this extremal background is then
\begin{align}
 \Delta M=\frac{\Omega}{16\pi G}(D-2)(\mu-\mu_e).\label{DeltaM_def}
\end{align}

The respective $U(1)$ charges are calculated by
\begin{align}
 Q_i&=\frac{1}{16\pi G}\oint\dif^{D-2} y\sqrt{\tilde{\gamma}}\expo{-2 a_i\psi}(F_i)_{\mu\nu}n^\mu\xi^\nu,
\end{align}
for $i=1$, $2$, and $3$. Calculating the Maxwell tensors from the potentials given in Eqs.~\Eqref{A1}, \Eqref{A2} and \Eqref{A3}, the charges are explicitly
\begin{align}
 Q_1&=\frac{\Omega}{16\pi G}\expo{-a_1\psi_0}\sqrt{2(n-\nu)(D-1+n-2\nu)}\;\frac{1}{L},\\
 Q_2&=\frac{\Omega}{16\pi G}\expo{-a_2\psi_0}\sqrt{2(D-2)(D-3+n)}\;q,\\
 Q_3&=\frac{\Omega}{16\pi G}\expo{-a_3\psi_0}\sqrt{2(D-3+n)}\rho.
\end{align}
We shall also fix the respective gauge potentials so that each  $A_i$ are zero at the horizon. This gives 
\begin{align}
 \Phi_1&=-\expo{\alpha_1\psi_0}\sqrt{\frac{2(n-\nu)}{D-1+n-2\nu}}\frac{r_+^{D-1+n-2\nu}}{L},\\
 \Phi_2&=\expo{\alpha_2\psi_0}\sqrt{\frac{2(D-2)}{D-3+n}}\;\frac{q}{r_+^{D-3+n}},\\
 \Phi_3&=-\expo{\alpha_3\psi_0}\sqrt{\frac{2}{D-3+n}}\rho r_+^{D-3+n}.
\end{align}

The horizon area, which will be crucial in the thermodynamic analysis in the next section, is 
\begin{align}
 \mathcal{A}&=r_+^{D-2}\Omega. \label{HorizonArea}
\end{align} 
Briefly looking at some curvature invariants, the Ricci and Kretschmann scalars are 
\begin{align}
 R&=-\frac{1}{r^n}\brac{\frac{f'}{r^n}}'+\frac{(D-2)}{r^{1+n}}\brac{\frac{2n}{r}f-2f'}+\frac{(D-3)(D-2)}{r}\brac{k-\frac{f}{r^n}},\label{RicciScalar}\\
 R_{\mu\nu\rho\sigma}R^{\mu\nu\rho\sigma}&=\frac{1}{r^{2n}}\sbrac{f''^2-\frac{2n}{r}f''f'+\frac{n^2}{r^2}f'^2}+\frac{D-2}{r^{2n+2}}f^2\brac{\frac{2f'^2}{f^2}+\frac{4n^2}{r^2}-\frac{4n}{r}\frac{f'}{f}}\nonumber\\
   &\quad+\frac{2(D-2)(D-3)}{r^4}\brac{k-\frac{1}{r^{2n}}f}^2, \label{KretschmannScalar}
\end{align}
we find that a curvature singularity occurs for $r=0$, and it persists in the `no black hole' case $\mu=0$, which is similar to the pure Lifshitz spacetime, as well as the zero mass limit of the dilaton black hole with Liouville potentials. Furthermore, we see that the curvature invariants are regular at the horizon $r=r_+$.


\section{Thermodynamics} \label{thermodynamics}

\subsection{The thermodynamic variables}

In this section, we shall consider the thermodynamic behaviour of the black hole in the case $M=N=3$. Similar to the Lifshitz black hole of \cite{Tarrio:2011de}, the first and third gauge fields, $A_1$ and $A_3$ are fixed by the asymptotic structure of the spacetime, and does not participate in the mechanics of the black hole. This is compounded by the fact that they are diverging at the boundary and may not have an appropriate holographic dual, if one were considering Lifshitz holography. 

We observe that, similar to the situation in Ref.~\cite{Tarrio:2011de}, the potentials $A_1$ and $A_3$ diverge at the boundary. Though the former is requied for a spherical horizon while the latter supports the asymptotic structure of the spacetime. These two fields, along with their associated conserved charges $Q_1$ and $Q_3$, does not affect the thermodynamics of the black hole.   In our present context, the condition for a spherical horizon $k=1$ is obtained by adjusting two parameters, namely $\rho$ and $\Lambda_3$ in Eq.~\Eqref{B}. Therefore it is possible to set $\rho=0$ by choosing an appropriate $\Lambda_3$. Hence $A_3$ vanishes altogether.  Thus, $A_1$, $A_2$, $Q_1$, and $Q_3$ do not play any role as thermodynamic variables. 

On the other hand, $A_2$ and its associated conserved charge $Q_2$ play the role of the Reissner--Nordstr\"{o}m-type potentials and charges, and does contribute to the thermodynamic behaviour of the solution. As they are the only $U(1)$ gauge fields relevant to the thermodynamic analysis, in this section we shall drop the subscripts and simply refer to them as $A$ and $Q$.

%

Additionally, as there are no conserved charges associated with the scalar field $\psi$, this field is not involved in the thermodynamics as well. Hence the scalar field is fixed throughout the thermodynamic analysis. In the thermodyanmic analsyis, the constant $\psi_0$ only modifies the values of $Q$ and $A$ by an overall factor. Hence may set $\psi_0=0$ without loss of generality. 

The temperature of the black hole can be obtained by ensuring the Euclidean time $\tau=\im t$ has the appropriate periodicity so that the Euclideanised metric is regular at the horizon \cite{Hawking:1980gf}. The temperature $T$ is then obtained as the inverse of this periodicity, where the resulting expression is $T=\frac{1}{4\pi}r_+^{-n}f'(r_+)$. In terms of the spacetime parameters, the temperature is explicitly
\begin{align}
 T&=\frac{1}{4\pi}\bigg[(D-1+n-2\nu)L^2r_+^{1+n-2\nu}+(D-1-n)\ell^{-2} r_+^{1-n}+(D-3+n)Br_+^{-1+n}\nonumber\\
    &\quad\hspace{4cm}-(D-3+n)q^2r_+^{-(2D-5+n)}\bigg].
\end{align}
The case of extremal black holes is where the temperature is zero. The value of horizon radius which gives $T=0$ is $r_+=r_e$, where $r_e$ satisfies the equation
\begin{align}
 q^2r_e^{-(D-3+n)}=\frac{D-1+n-2\nu}{D-3+n}L^{-2}r_e^{D-1+n-2\nu}+\frac{D-1-n}{D-3+n}\ell^{-2}r_e^{D-1-n}+Br_e^{D-3+n}.
\end{align}
The corresponding mass parameter in the extremal case is
\begin{align}
 \mu_e=\frac{2(D-2+n-\nu)}{D-3+n}L^{-2}r_e^{D-1+n-2\nu}+\frac{2(D-2)}{D-3+n}\ell^{-2}r_e^{D-1-n}+2Br_e^{D-3+n}.
\end{align}
The entropy of the black hole is, in the units of the present paper, is 
\begin{align}
 S=\frac{\mathcal{A}}{4G},
\end{align}
where $\mathcal{A}$ is the horizon area given by Eq.~\Eqref{HorizonArea}.

%

In the grand canonical ensemble, the gauge potential is held fixed at the boundary at value $\Phi$, and serves as the variable conjugate to the charge $Q$,
\begin{align}
 \Phi&=\sqrt{\frac{2(D-2)}{D-3+n}}qr_+^{-(D-3+n)}.
\end{align}
The energy $E$ in this ensemble is taken to be the black hole mass $M$ as given by \Eqref{M_def}. With these quantities, it can be checked that the first law of thermodynamics hold: 
\begin{align}
 \dif E&=T\dif S+\Phi\dif Q.
\end{align}

For the canonical ensemble, the charge $Q$ is held fixed. Therefore $\dif Q=0$ and we take its conjugate variable to be the difference of the potential with the extremal case, $\Phi-\Phi_e$, where $\Phi_e=\sqrt{\frac{D-2}{2(D-3+n)}}qr_e^{-(D-3+n)}$. Also, the energy should appropriately be measured against the background of the extremal spacetime, where its horizon has zero temperature. Hence we shall take the energy of the canonical ensemble to be $\Delta E=\Delta M$, where $\Delta M$ is given in Eq.~\Eqref{DeltaM_def}. In this case, the first law can be checked to read 
\begin{align}
 \dif\brac{\Delta E}=T\dif S.
\end{align}

\subsection{Canonical ensemble}

The relevant thermodynamic potential in the canonical ensemble is the Helmholtz free energy 
\begin{align}
 \mathcal{F}&=\Delta E-TS\nonumber\\
            &=\frac{\Omega}{16\pi G}\bigg[-(1+n-2\nu)\ell^{-2}r_+^{D-1+n-2\nu}-(1-n)\ell^{-2}r_+^{D-1-n}\nonumber\\
            &\quad\hspace{1.5cm}+(1-n)Br_+^{D-3+n}+(2D-5+n)q^2r_+^{-(D-3+n)}\nonumber\\
            &\quad\hspace{0.2cm}-(D-2)\brac{\frac{D-1+n-2\nu}{D-3+n}L^{-2}r_e^{D-1+n-2\nu}+\frac{D-1-n}{D-3+n}\ell^{-2}r_e^{D-1-n}+Br_e^{D-3+n}}\bigg].
\end{align}
Let us also write the temperature here explicitly for convenient reference:
\begin{align}
 T&=\frac{1}{4\pi} \bigg[(D-1+n-2\nu)L^{-2}r_+^{1+n-2\nu}+(D-1-n)\ell^{-2}r_+^{1-n}+(D-3+n)Br_+^{-1+n}\nonumber\\
    &\quad\hspace{4cm}-(D-3+n)q^2r_+^{-(2D-5+n)}\bigg].\label{T_can}
\end{align}
The heat capacity at constant charge is
\begin{align}
 C_Q&=T\brac{\frac{\partial S}{\partial T}}_Q=\frac{\partial M/\partial r_+}{\partial T/\partial r_+},
\end{align}
where
\begin{align}
 \frac{\partial M}{\partial r_+}&=\frac{\Omega(D-2)}{4G}r_+^{D-3}T,\\
 \frac{\partial T}{\partial r_+}&=\frac{1}{4\pi}r_+^{-1}\Big[(1+n-2\nu)(D-1+n-2\nu)L^{-2}r_+^{1+n-2\nu}+(1-n)(D-1-n)\ell^{-2}r_+^{1-n}\nonumber\\
 &\quad\quad -(1-n)(D-3+n)Br_+^{-(1-n)}+(2D-5+n)(D-3+n)q^2r_+^{-(2D-5+n)}\Big].
\end{align}

Recall that for fixed $n$, the parameter $\nu$ ranges from $\nu=0$ to $\nu=n$. The former case being the Lifshitz black hole and the latter case corresponds to a charged dilaton black hole with two exponential potentials. We find that both cases share similar thermodynamic behaviour connected by the continuous parameter $\nu$. 

In particular, for sufficiently small charge, there exist a possible range where three branches of solutions with different $r_+$ share the same temperature. This is reflected in Eq.~\Eqref{T_can} where the equation $T_0=T(r_+)$ has possibly multiple roots for some $T_0$. As $q$ is increased until $q_{\mathrm{crit}}$, the roots will coalesce to a single point, after which for $q>q_{\mathrm{crit}}$ there will be a unique $r_+$ for every $T$. Given some spacetime parameters $(n,\nu,L^2,\ell^{2},B)$ at some dimensionality $D$, the value of $q_{\mathrm{crit}}$ is determined by the condition 
\begin{align}
 \frac{\partial T}{\partial r_+}=\frac{\partial^2T}{\partial r_+^2}=0. \label{q_crit_eqn}
\end{align}

\begin{figure}
 \begin{center}
  \includegraphics{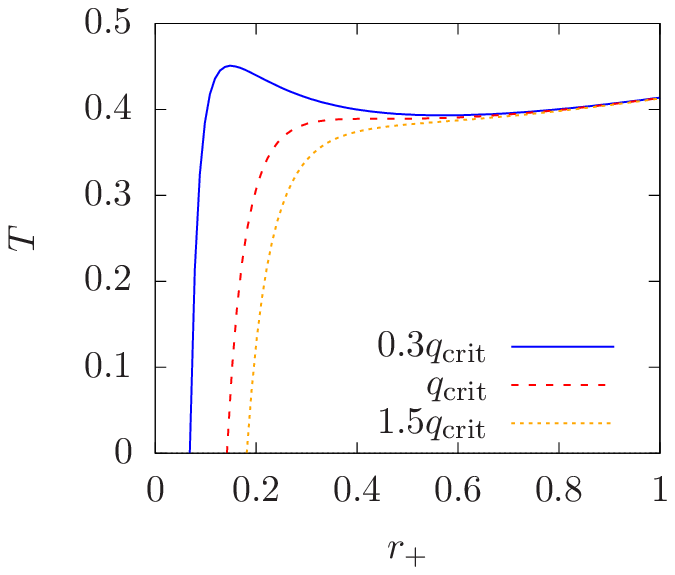}
  \includegraphics{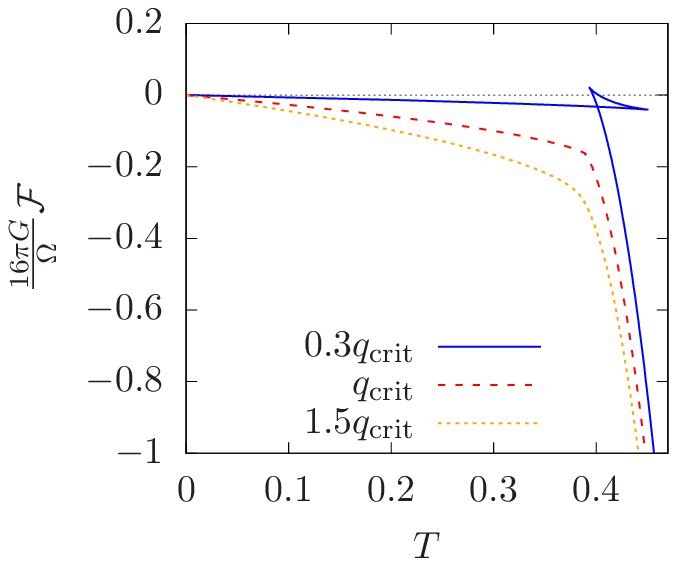}
  \caption{(Colour online) Plots of $T$ vs $r_+$ (left) and $\mathcal{F}$ vs $T$ (right) for the case $n=0.6$, $\nu=0.3$, $L^2=\ell^2=2$, and $D=4$. In this case, the critical charge has the value $q_{\mathrm{crit}}=0.0588$.}
  \label{fig_01}
 \end{center}

\end{figure}

Fig.~\ref{fig_01} shows a representative example for the case $n=0.6$, $L^2=2$, $\ell^2=2$, and $D=4$. In this case, solving Eq.~\Eqref{q_crit_eqn} gives $q_{\mathrm{crit}}=0.0588$. For a charge less than $q_{\mathrm{crit}}$, there are three co-existing branches of black holes with temperature $0.3931\leq T\leq 0.4508$, which can be seen in in the solid curve in the left-hand plot in Fig.~\ref{fig_01}. 
The plot of $\mathcal{F}$ vs $T$ for this charge is shown in the solid curve on the right-hand plot of Fig.~\ref{fig_01}, where we can see it's characteristic swallowtail structure. Recall that the thermodynamically favoured phase is the one with lower Helmholtz free energy. Hence for a given $T$ with multiple values of $\mathcal{F}$, the lower branch is preferred. At $q=q_{\mathrm{crit}}$, the three branches collapse to a point and the swallowtail in the $\mathcal{F}$-$T$ curve in shrinks to a single kink. Further increasing beyond $q>q_{\mathrm{crit}}$, the kink disappears and $\mathcal{F}$ becomes a smooth function of $T$. 


\begin{figure}
 \begin{center}
  \includegraphics{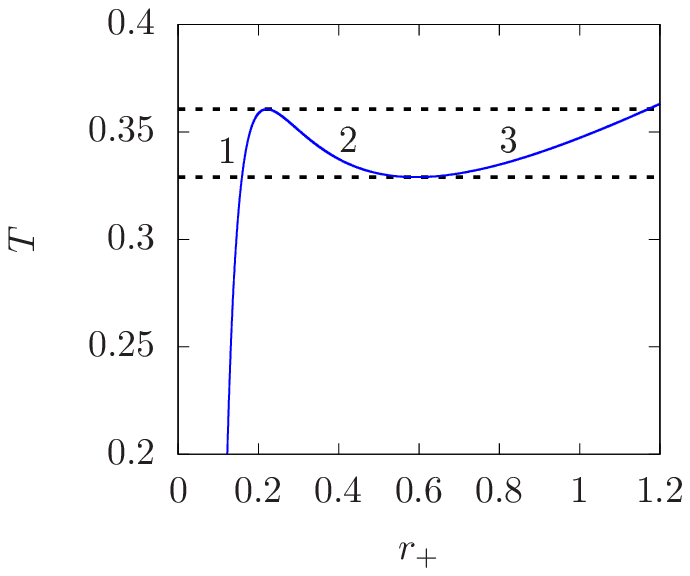}
  \includegraphics{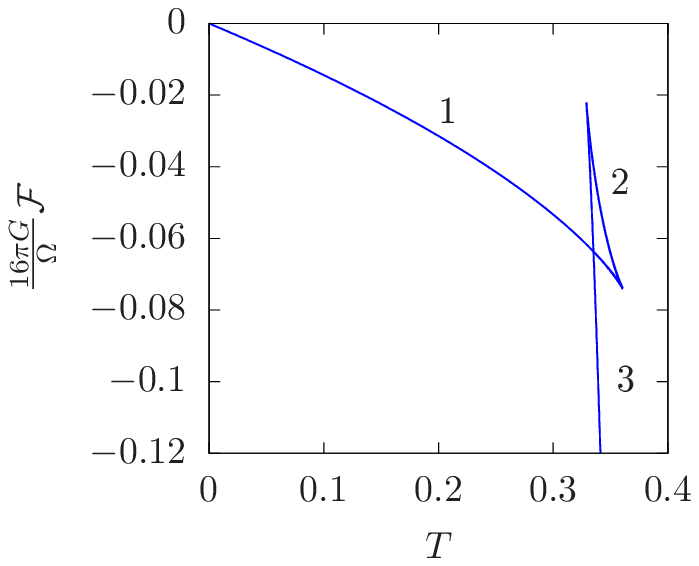}
  \caption{(Colour online) Plots of $T$ vs $r_+$ (left) and $\mathcal{F}$ vs $T$ (right) for $q=\sqrt{0.3}q_{\mathrm{crit}}$ in the case $n=0.4$, $\nu=0.3$, $L^2=\ell^2=2$, and $D=4$. In this case, the critical charge has the value $q_{\mathrm{crit}}=0.0868$.}
  \label{fig_02}
 \end{center}

\end{figure}

Let us explore another case of $q<q_{\mathrm{crit}}$ in further detail, this time with $n=0.4$, $\nu=0.3$, $L^2=\ell^2=2$ and $D=4$. The three branches are labelled explicitly in Fig.~\ref{fig_02}. Branch 1 is a low-temperature solution which starting from the extremal case $r_+=r_e$ up to $r_+=0.2209$. wheras Branch 2 is an unstable one at $0.2209< r_+ < 0.5920$, and Branch 3 is for $r_+> 0.5920$. From Fig.~\ref{fig_02}, we see that the three branches can co-exist in the temperature range $0.3290\leq T\leq 0.3606$, and that a first-order phase transition may occur in this range \cite{Tarrio:2011de,Chamblin:1999tk}.

\begin{figure}
 \includegraphics[scale=0.95]{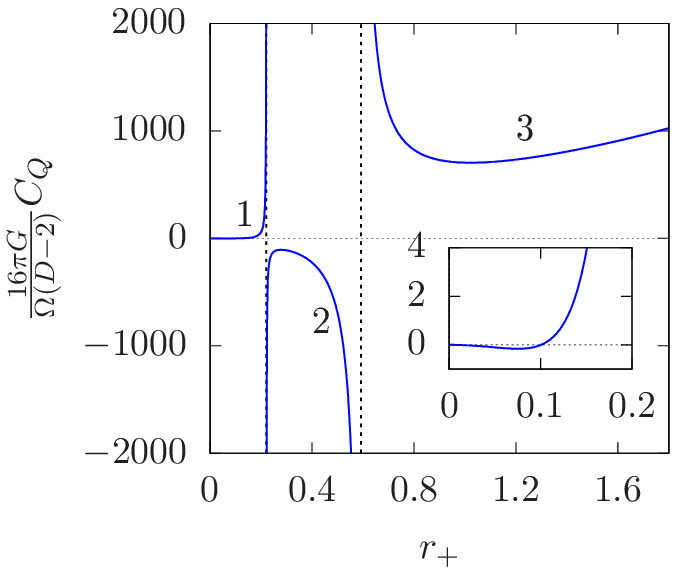}
 \includegraphics[scale=0.95]{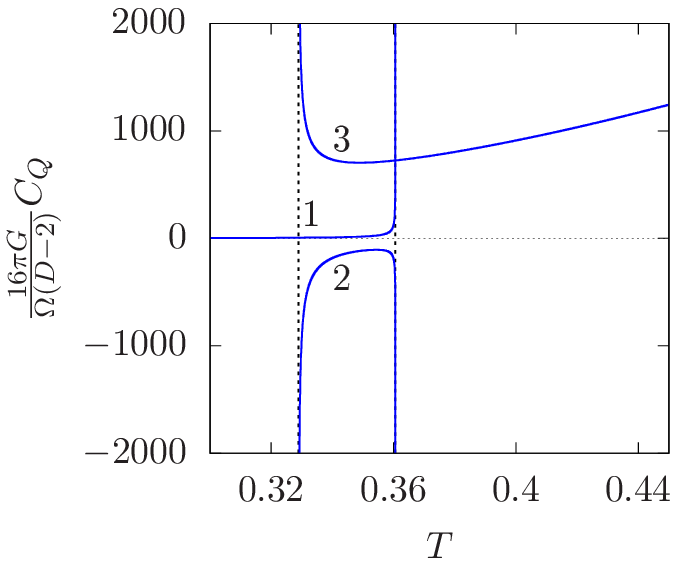}
 \caption{(Colour online) Plots of $C_Q$ vs $r_+$ (left) and $C_Q$ vs $T$ (right). The parameter choices are the same as for Fig.~\ref{fig_02}.}
 \label{fig_03}
\end{figure}

This behaviour is further corroborated by the heat capacity. The left-hand plot of Fig.~\ref{fig_03} shows the heat capacity $C_Q$ vs $r_+$. The heat capacities of Branches 1 and 3 are positive while it is negative for Branch 2, and they are separated by discontinuities. The plot of $C_Q$ vs $T$ is shown in the right-hand plot of Fig.~\ref{fig_03}, where we observe the co-existing branches in the aforementioned temperature range. 

\begin{figure}
 \begin{center}
  \includegraphics[scale=1]{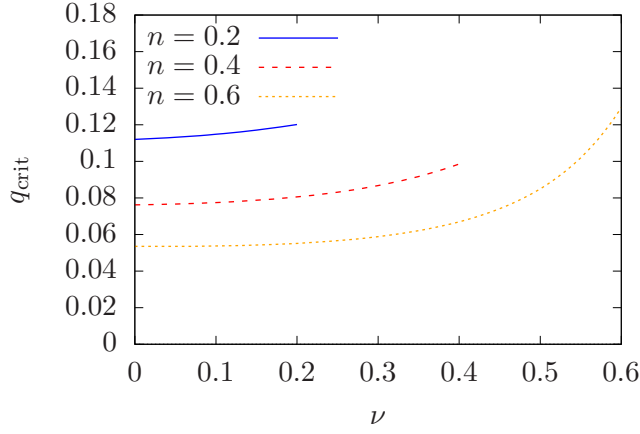}
  \caption{(Colour online) Plots of $q_{\mathrm{crit}}$ against $\nu$ for various $n$. Note that the allowed ranges of $\nu$ are $\nu\leq n$, and hence each curve terminates at their respective $\nu=n$.}
  \label{fig_nuvsqcrit}
 \end{center}

\end{figure}

The thermodynamics typically show similar qualitative behaviour for $0\leq\nu\leq n$, where $\nu$ determines the specific value of $q_{\mathrm{crit}}$.  
In Fig.~\ref{fig_nuvsqcrit}, some values of $q_{\mathrm{crit}}$ as a function of $\nu$ are shown for different $n$. Furthermore, as $\nu$ ranges from $\nu=0$ to $\nu=n$, these observations connect the thermodynamics of the Lifshitz black hole \cite{Tarrio:2011de} to that of a charged dilaton black hole with two Liouville potentials \cite{Sheykhi:2007wg,Hendi:2015xya}.

\subsection{Grand canonical ensemble} 

\begin{figure}
 \includegraphics{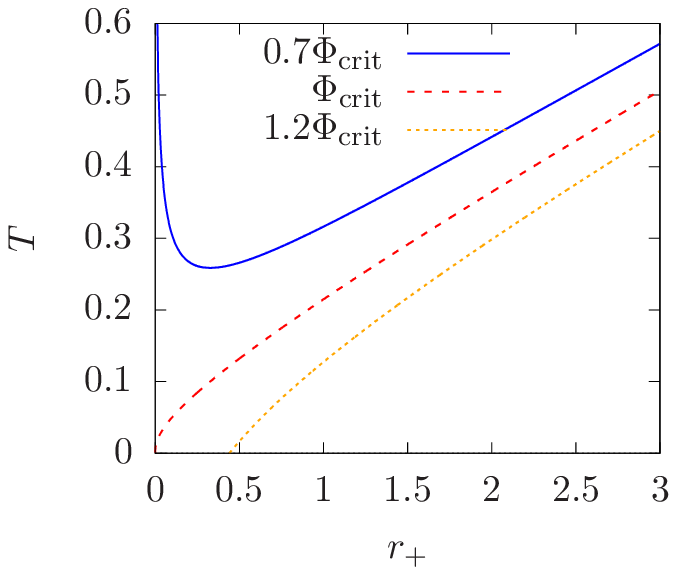}
 \includegraphics{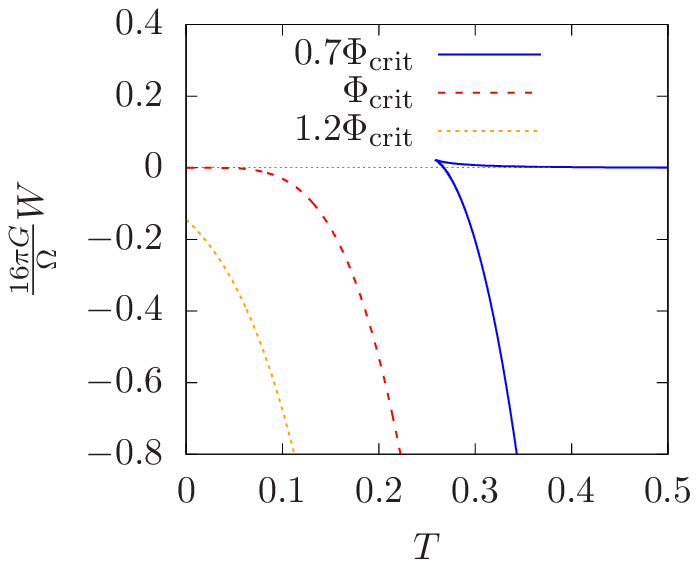}
 \caption{Plots of $T$ vs $r_+$ (left) and $W$ vs $T$ for the case $n=0.6$, $\nu=0.3$, $L^2=\ell^2=2$, and $D=4$. For these values, the critical potential is $\Phi_c=0.9882$.}
 \label{fig_04}
\end{figure}

In the grand canonical ensemble, the gauge potential $\Phi$ is held fixed instead of $Q$. In this case, the relevant thermodynamic potential is the Gibbs free energy 
\begin{align}
 W&=E-TS-\Phi Q\nonumber\\
  &=\frac{\Omega}{16\pi G}\bigg[-(1+n-2\nu)L^{-2}r_+^{D-1+n-2\nu}-(1-n)\ell^{-2}r_+^{D-1-n}\nonumber\\
  &\quad\hspace{2cm}+(1-n)\brac{B-\frac{2(D-3+n)}{D-2}\Phi^2}r_+^{D-3+n}\bigg].
\end{align}
In this ensemble, the system should be parametrised by $\Phi$ instead of $q$. Therefore the temperature is expressed as 
\begin{align}
 T&=\frac{1}{4\pi}\bigg[(D-1+n-2\nu)L^{-2}r_+^{1+n-2\nu}+(D-1-n)\ell^{-2}r_+^{1-n}\nonumber\\
  &\quad\hspace{2cm} +(D-3+n)\brac{B-\frac{2(D-3+n)}{D-2}\Phi^2}r_+^{-1+n}\bigg].\label{Tgc}
\end{align}
The above relations imply the existence of a critical potential $\Phi_{\mathrm{crit}}$, where 
\begin{align}
 \Phi_{\mathrm{crit}}=\sqrt{\frac{(D-2)B}{2(D-3+n)}}.
\end{align}
For $\Phi<\Phi_{\mathrm{crit}}$, the third term in Eq.~\Eqref{Tgc} is negative, and this allows the possibility two branches of solution with the same temperature. the free energies of these two branches meet at a discontinuous point of the $W$-$T$ curve, as shown in Fig.~\ref{fig_04}, where the numerical values are $n=0.6$, $\nu=0.3$, $L^2=\ell^2=2$, and $D=4$. Increasing the potential to $\Phi_{\mathrm{crit}}$, we see that the low-temperature branch vanishes by collapsing into the point $(r_+,T)=(0,0)$ and only one smooth branch $W$-$T$ curve remains. This behaviour presists for various values of $\nu$ in $0\leq \nu\leq n$, and is similar to the thermodynamics observed in the Riessner--Nordstr\"{o}m--Anti-de Sitter black hole \cite{Chamblin:1999tk} as well as the spherical Lifshitz black hole \cite{Tarrio:2011de}.   

\section{Conclusion} \label{conclusion}

In summary, we have considered an Einstein--Maxwell--dilaton-type theory consisting of $M$ Liouville potentials and $N$ $U(1)$ gauge fields. With an appropriate ansatz, the equations of motion are solved for arbitrary $M$ and $N$.   

The ansatz was chosen based on the assumption that each term in the metric function $g_{tt}$ is supported by a potential and gauge field, allowing for the possibility that a potential and gauge field may support the same term. By switching off the appropriate potentials and gauge fields, we recover the Lifshitz and dilaton black holes. The solution with $M=N=3$ corresponds to a metric that interpolates between the two. 

Some basic thermodynamic quantities have been studied. In the canonical ensemble, solutions with sufficiently small charge may have three co-existing branches with the same temperature, and that a phase transition may occur among them. In the grand canonical ensemble, solutions with sufficiently small potential, there are two branches with the same temperature. This behaviour shares similar features to the Lifshitz black hole as well as the dilaton black hole to which it interpolates, thus bridging the results of \cite{Tarrio:2011de} for the Lifshitz black hole and \cite{Sheykhi:2007wg,Hendi:2015xya} for the dilaton black hole. It might be interesting to find exact solutions for other forms of the dilaton potential $\Vcal$, as was shown in \cite{Astefanesei:2019mds} for asymptotically-AdS solutions, there are certain forms of potential that lead to thermodynamically stable exact solutions.

Our solution inherits similar problematic features of the Lifshitz and dilaton black holes. In particular, the dilaton $\psi$ diverges at the boundary. It was argued in \cite{Tarrio:2011de} that this issue might be resolved by an appropriate embedding of the model in string theory. Secondly, the unsual asymptotic structure, being non-asymptotically flat and non-(A)dS complicates the issue of calculations at the boundary. For instance, the choice of background required to renormalise the divergence of the Euclidean action is ambiguous. Instead, a background-independent counter-term was introduced in the Lifshitz case \cite{Tarrio:2012xx} and similarly for the dilaton black hole \cite{Cai:1999xg}. However, in the dilaton case, the boundary stress tensor remains ill-defined despite having a finite action \cite{Cai:1999xg}. Since the solution of the present paper contains the dilaton black hole as a special case, it inherits the similar issue of the dilaton black hole's boundary stress tensor.


\section*{Acknowledgements}
This research is supported by Xiamen University Malaysia Research Fund (Grant no. XMUMRF/2019-C3/IMAT/0007).

\bibliographystyle{mnbh}

\bibliography{mnbh}

\end{document}